\documentclass[prd,preprint,superscriptaddress,nofootinbib]{revtex4}

\usepackage{amsmath,amssymb,array,calc,epsfig}
\usepackage{graphicx}

\newlength{\intwidth}

\def\sk{k\!\!\!\slash}
\def\h{\hbox{-}}

\begin{document}

\title{Sharpening The Leading Singularity}

\author{Freddy Cachazo}
\email{fcachazo@perimeterinstitute.ca}

\affiliation{Perimeter Institute for Theoretical Physics, Waterloo,
Ontario N2J 2W9, Canada}

\begin{abstract}

We show how studying leading singularities of Feynman diagrams, when
all momenta are complex, gives a simple way of writing multi-loop
and multi-particle scattering amplitudes in ${\cal N}=4$ super
Yang-Mills. The simplicity of the method is equivalent to that of
the quadruple cut technique introduced in hep-th/0412103 at
one-loop. The new technique only involves the computation of
residues and the solution of linear equations. In our technique both
parity even and parity odd pieces of a coefficient are computed
simultaneously and it is only at the end that a separation can be
made if desired. We explain the procedure via examples. The main
example, which we compute in detail, is the five-particle two-loop
amplitude first given in hep-th/0604074. Another feature of our
method is that the helicity structure of the amplitude only enters
in the inhomogeneous part of the linear equations. In other words,
the homogeneous part is universal. We illustrate this feature by
presenting the linear equations which determine a large class of
terms for MHV and next-to-MHV six-particle two-loop amplitudes.

\end{abstract}

\maketitle

\section{Introduction}

The idea of determining the $S$-matrix of a theory from the
structure of its singularities was intensively studied in the 60's
\cite{Weinberg:1964ev, Weinberg:1964ew,Weinberg:1965rz,Olive:1964,
Chew:1966, Olive:1967}. The main idea was to consider the $S$-matrix
as an analytic function of the kinematical invariants with
singularities determined by physical input. Most efforts were
directed towards the study of hadronic interactions. Nowadays we
know that understanding analytically the strong interactions is a
very hard problem.

A more modern version, similar in spirit, was developed in the 90's,
where extensive use of branch cut singularities was shown to be a
powerful tool in the computation of scattering amplitudes in
Yang-Mills theories. These techniques came to be known as the
unitarity based method
\cite{Bern:1994zx,Bern:1994cg,Bern:1995db,Bern:1996je,Bern:1996ja,Bern:1997sc,Bern:2004cz}.

More recently \cite{Britto:2004nc,Buchbinder:2005wp,Cachazo:2008dx}
, the highest codimension singularities, called leading
singularities already in the 60's, were shown to be a powerful tool
in computing amplitudes. At one-loop, the problem of computing
amplitudes is reduced to that of computing tree amplitudes for
theories where no triangles or bubbles appear. One such theory is
${\cal N}=4$ super Yang-Mills (SYM) and it has been hypothesized
that so too is ${\cal N}=8$ supergravity
\cite{Bern:1998sv,Bern:2005bb,BjerrumBohr:2005xx,BjerrumBohr:2006yw}.

In this paper we show that at higher loop orders the power of the
leading singularity has only been partially
unleashed~\cite{Buchbinder:2005wp,Cachazo:2008dx}. By carefully
studying all leading singularities we propose a method which reduces
the computation of multi-loop and multi-particle amplitudes in
${\cal N}=4$ SYM to the computation of residues (which end up being
related to tree amplitudes) and to the solution of {\it linear}
equations.

The discontinuity across the leading singularity at one loop is
computed by collecting all Feynman diagrams that share four given
propagators and then cutting the propagators to turn the sum into
simply the product of four tree-level amplitudes. Cutting means
removing the principal value part of a Feynman propagator, {\it
i.e.} $1/(\ell^2+i\epsilon)\rightarrow \delta^{(+)}(\ell^2)$. Thus
the integral over the four-vector $\ell$ is completely localized by
the four delta functions. The value of the integral for each
solution $\ell_*$ is given by the jacobian of the change of
variables from $\ell$ to the argument of the delta functions
evaluated at $\ell_*$. The final answer is then the sum over of all
contributions. In massless theories there are two solutions
$\ell_*^{(1)}$ and $\ell_*^{(2)}$. In general, the product of the
four tree-level amplitudes gives different answers at the different
points $\{ \ell_*^{(1)},\ell_*^{(2)}\}$. An important exception is
when the number of particles is four. In this case, the product of
amplitudes is equal to $A^{\rm tree}st$ for both values of $\ell$.

In general, the support of the delta functions is outside the region
where $\ell$ is a real vector. This means that the computation of
the leading singularity\footnote{Here and throughout the paper we
will abuse terminology and refer to the discontinuity across a given
leading singularity as the leading singularity itself.} is more
naturally interpreted in terms of a contour integral in $\Bbb{C}^4$
where $\ell$ is now a complex vector. There are two distinct leading
singularities and the prescription of~\cite{Britto:2004nc} is
equivalent to choosing a contour that picks up both residues. A
natural question is what the role of each isolated leading
singularity is. Naively, one could try to expand the one-loop
amplitude in scalar boxes and then compute their coefficients by
matching the residues at a given leading singularity. If one did
that one would find different answers for the same coefficient which
would be a contradiction.

In this paper we show that the resolution to this puzzle is that at
one-loop, reduction formulas that express {\it e.g.} pentagons in
terms of boxes, derived in \cite{Passarino:1978jh,van
Neerven:1983vr,Bern:1993kr}, are not valid for generic complex
integration contours. This means that in order to write an
expression in terms of scalar integrals which reproduces {\it all}
singularities of Feynman diagrams correctly one has to allow for
higher point integrals. This explains why four particles is special
and why the first non-trivial case is that of five particles, where
one should allow for a pentagon.

At higher loops, the original approach of~\cite{Buchbinder:2005wp}
is to sum over all solutions. In~\cite{Cachazo:2008dx}, it was
stated that one should be able to work with individual
singularities, but this was in the context of four-particles where
there is no distinction.

In this paper we show that requiring agreement between Feynman
diagrams and scalar (or generalized scalar) integrals on each
individual leading singularity provides enough linear equations to
determine higher-loop and multi-particle amplitudes.

In an nutshell, one starts collecting all Feynman diagrams with
chosen propagators that give rise to topologies with only boxes. By
writing an ansatz of scalar integrals ({\it i.e.}, with numerator
one) one requires agreement at all leading singularities
independently. This gives a set of linear equations. Sometimes the
system of equations does not have solutions which means that
generalized scalar integrals ({\it i.e.}, with numerators that are
propagators with negative power) must be added. The numerators serve
as zeroes to cancel poles so that the new integrals have zero
residue on leading singularities where they are not needed.

We explain the method via examples. Our main example is the
five-particle two-loop amplitude. This amplitude was first computed
in \cite{Bern:2006vw}. In this case, our computation only requires
solving linear equations in two variables!

An interesting new feature already found in \cite{Bern:1997nh}, as
compared to the four-particle amplitude, is that after normalizing
by the corresponding tree-level amplitude one finds terms that are
parity even and some that are parity odd. Using our technique, it
becomes very transparent why the answer does not have definite
parity. The reason is that leading singularities come in pairs. For
real external momenta, one of them would be located at the complex
conjugate value of the other. For four particles, the value of
Feynman diagrams on each leading singularity is the same and hence
one gets a parity even answer. For five or more particles, the
values differ and one gets complex solutions.

It is important to mention that the actual helicity structure of the
amplitude only enters as the inhomogeneous part of the linear
equations that determine the coefficients of the integrals. In this
sense, the homogeneous part of the linear system of equations is
universal. Since for five particles, all non-trivial helicity
configurations are MHV (or ${\overline{\rm MHV}}$) the feature just
described is not very surprising. This is why we present the linear
equations which determine a large class of terms contributing to
six-particle MHV and next-to-MHV two-loop amplitudes. Solving the
equations could allow a comparison to the recent computation of the
parity even part of the six-particle MHV amplitude
in~\cite{Bern:2008ap}.

This paper is organized as follows. In section II we establish our
conventions. In section III we give a detailed explanation of the
leading singularity technique at one-loop. The discussion of how the
scalar basis must be extended is given for five particles since the
results there can be directly applied to the two-loop case. In
section IV we introduce the leading singularity technique at two
loops in the simplest case, {\it i.e.}, the four-particle amplitude.
In section V we present the main example of the paper, the
five-particle two-loop amplitude, in complete detail. In section VI
we present the linear equations for coefficients in MHV and
next-to-MHV six-particle two-loop amplitudes. In section VII we give
conclusions and future directions.

\section{Preliminaries}
\label{sec:prelims}

Scattering amplitudes of on-shell particles in ${\cal N}=4$ SYM with
gauge group $U(N)$ can be written as a sum over color-stripped
partial amplitudes using the color
decomposition~\cite{Bern:1996je,Mangano:1990by,Dixon:1996wi}. Each
partial amplitude admits a large $N$ expansion. More explicitly,
\begin{eqnarray}
{\cal A}_n(1,2,\ldots , n) &=& \delta^{(4)}(p_1+p_2+\ldots + p_n)\
{\rm
Tr}(T^{a_1}T^{a_2}\ldots T^{a_n})\ A_n(1,2,\ldots, n) \nonumber\\
&&\qquad\qquad + \ {\rm permutations}\ +\ \ldots
\end{eqnarray}
where the sum is over non-cyclic permutations of the external states
(cyclic ones being symmetries of the trace) and the ellipsis
represents terms with double and higher numbers of traces. $A_n$ may
be expanded in perturbation theory and we denote the $L$-loop planar
partial amplitude by $A_n^{(L)}$. We also use $A_n^{(0)}=A_n^{\rm
tree}$.

Our conventions are:

\begin{itemize}

\item A tree-level MHV amplitude has two particles of negative
helicity and the rest of positive helicity.

\item A null vector $p_\mu$ is written as a bispinor as $p_{a\dot
a}=\lambda_a\tilde\lambda_{\dot a}$.

\item The Lorentz invariant inner product of two null vectors $p_\mu$ and $q_\mu$ is given by $2p\cdot q =
\langle \lambda_p, \lambda_q\rangle
[\tilde\lambda_p,\tilde\lambda_q]$. We also use $\langle \lambda_p,
\lambda_q\rangle = \langle p,q\rangle$ and $
[\tilde\lambda_p,\tilde\lambda_q] = [p,q]$.

\item All external momenta in an amplitude are outgoing and usually
denoted by $k_i$.

\item Some useful Lorentz invariant combinations are
$s_{ij}=(k_i+k_j)^2$, $t_{ijl}=(k_i+k_j+k_l)^2$, and
$[a|b+c|d\rangle = [a,b]\langle b,d\rangle+[a,c]\langle c,d\rangle$.

\end{itemize}

\section{Leading Singularity at One-Loop}
\label{sec:1loop}

Any one-loop amplitude in ${\cal N}=4$ SYM may be expressed both as
a sum over Feynman diagrams, and also in terms of scalar box
integrals \cite{Bern:1994zx}:
\begin{equation}
A_n^{(1)}= \sum\left\{\hbox{1-loop Feynman diagrams}\right\} =
\sum_{\cal I} B_{\cal I}\times I(K_1^{\cal I},K_2^{\cal I},K_3^{\cal
I},K_4^{\cal I}) \label{eq:quadruple}
\end{equation}
where the second sum is over all partitions ${\cal I}$ of
$\{1,2,\ldots, n\}$ into four non-empty sets, $K_i^{\cal I}$ equals
the sum of the momenta in the $i^{\rm th}$ subset of partition
${\cal I}$ and the $B_{\cal I}$ are coefficients to be determined.
Since we are working with color ordered Feynman diagrams, one
considers only partitions that respect the color ordering.

Scalar box integrals are of the form
\begin{equation}
I(K_1,K_2,K_3,K_4) := \int\frac{d^4\ell}{(2\pi)^4}
\frac{1}{\ell^2(\ell-K_1)^2(\ell-K_1-K_2)^2(\ell+K_4)^2}\ .
\label{eq:box}
\end{equation}

Since scalar integrals are known explicitly (see {\it e.g.}
\cite{smirnov}), the problem of computing any one-loop amplitude is
reduced to that of determining the coefficients $B_{\cal I}$.

The amplitude, defined in terms of Feynman diagrams, possesses many
singularities. Some of them are branch cut singularities and
comparing them on both sides is a way of obtaining information about
the coefficients and in many cases it allows their determination
like for MHV amplitudes \cite{Bern:1994zx}. The main difficulty is
that a given branch cut is generically shared by several scalar
integrals and therefore several coefficients appear at the same time
and the information must be disentangled.

\begin{figure}
\includegraphics[scale=0.50]{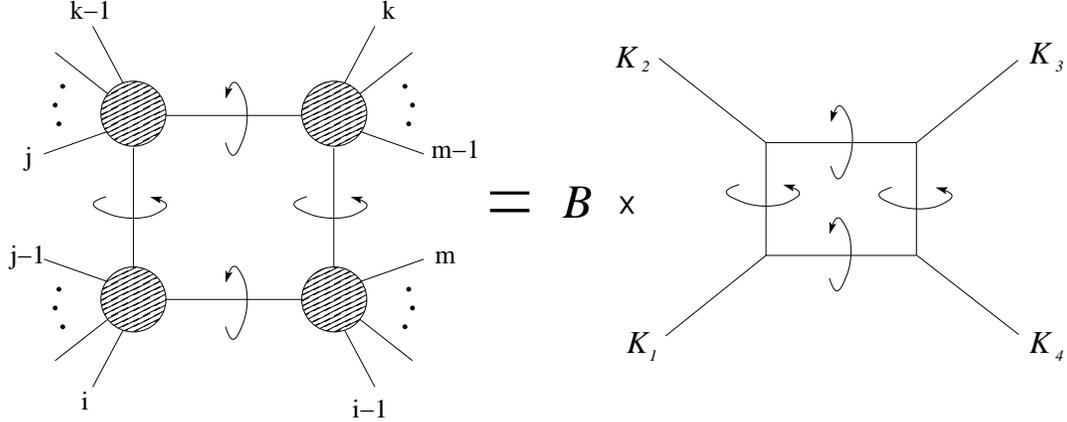}
\caption{Computation of a coefficient using the leading singularity
of a box. The lines circling the propagators represent the $T^4$
contour of integration. The left hand side of the figure represents
the sum of all 1-loop Feynman diagrams - note that only those
Feynman diagrams that contain the displayed propagators actually
contribute to this particular contour integral.}
\label{fig:quadruple}
\end{figure}

It turns out that the highest codimension singularities of Feynman
diagrams, called leading singularities, receive contributions from a
{\it single} scalar box integral. Thus, using leading singularities
of Feynman diagrams one can determine all coefficients, $B_{\cal
I}$, one by one \cite{Britto:2004nc}.

If we let $f_i(\ell)$ with $i=1,\ldots, 4$ correspond to the four
factors in the denominator of~(\ref{eq:box}), then the leading
singularity is computed by replacing $1/f_i(\ell)$ by
$\delta(f_i(\ell))$. Applying this to both sides
of~(\ref{eq:quadruple}) one finds that
\begin{equation}
\sum_{\ell_*} \left.\det\left(\frac{\partial
f_i}{\partial\ell_\mu}\right)^{-1}\sum_{\rm Multiplet}\prod_{i=1}^4
A^{{\rm tree}\ (i)}\right|_{\ell=\ell_*} = \left.B_{\cal J}\times
\sum_{\ell_*}\det\left(\frac{\partial
f_i}{\partial\ell_\mu}\right)^{-1}\right|_{\ell=\ell_*}\ ,
 \label{eq:wipi}
\end{equation}
where the sum over $\ell_*$ means a sum over all solutions to the
equations $f_i(\ell)=0$ for $i=1,\ldots,4$. In general there are two
solutions. The second sum on the l.h.s. is over all members of the
${\cal N}=4$ supermultiplet as choices of internal particles.

When the number of external particles is four, {\it i.e.} $n=4$,
both solutions give a jacobian factor equal to $1/(st)$ and the sum
over the product of tree-level amplitudes equals $A^{\rm tree}st$.
This implies that $B=A^{\rm tree}st$. For five or more particles the
sum over tree amplitudes is not the same for both solutions. In
fact, for five particles the product of tree amplitudes vanishes in
one solution and it gives $A^{\rm tree}st$ in the other. Here
$s=(K_1+K_2)^2$ and $t=(K_2+K_3)^2$ with $K_i$'s as
in~(\ref{eq:box}). Using this in~(\ref{eq:wipi}) one finds that $B
=A^{\rm tree}st/2$ which is the correct answer~\cite{Dixon:1996wi}.
Note the factor of half coming from the fact that on the r.h.s.
of~(\ref{eq:wipi}) one gets a factor of two.

\subsection{Sharpening The Leading Singularity}

The solutions, $\ell_*$, to the equations $f_i(\ell)=0$ are complex
in general. This makes the mathematical interpretation of the
leading singularity more transparent if defined as a contour
integral in $\Bbb{C}^4$. The contour integral is obtained by simply
taking~(\ref{eq:box}) and allowing $\ell$ to be a complex vector.
The contour of integration has the topology of a four-torus,
$T^4\cong (S^1)^4$, around each isolated singularity, and it is
given by $\Gamma = \left\{\ell : |f_i(\ell)|=\epsilon_i\right\}$
with $\epsilon_i$ some small positive number. The integral is
computed by a generalization of Cauchy's theorem to higher
dimensions (see for example section 5.1 of \cite{GH}). This can be
seen by a change of variables to local coordinates $z_i$'s in terms
of which the singular point is located at the origin, {\it i.e.}
$z_i= f_i(\ell)=0$. Now it is clear that the contour of integration,
$\Gamma = \left\{\ell : |z_i|=\epsilon_i\right\}$, when projected on
each $z_i$-plane is just a circle of radius $\epsilon_i$. The
integral becomes\footnote{Here and throughout the paper we use the
convention that the symbol $\oint$ contains a factor of $1/(2\pi
i)$.}
\begin{equation}
I = \left(\prod_{i=1}^4\oint_{z_i=0} \frac{dz_i}{z_i}\right)
\det\left(\frac{\partial z_i}{\partial\ell_\mu}\right)^{-1}.
\end{equation}

A natural questions that comes from this formalism is how to compare
the sum over Feynman diagrams and the sum over scalar integrals when
evaluated on each leading singularity independently. In other words,
we would like to make sense of~(\ref{eq:wipi}) term by term in the
sum over solutions. The reader might anticipate a problem from the
fact that the sum over tree amplitudes gives different answers for
$n>4$. Before giving the resolution to the puzzle explicitly for
$n=5$ let us discuss in detail the four-particle case.

\subsubsection{Four Particles}

\begin{figure}
\includegraphics[scale=0.50]{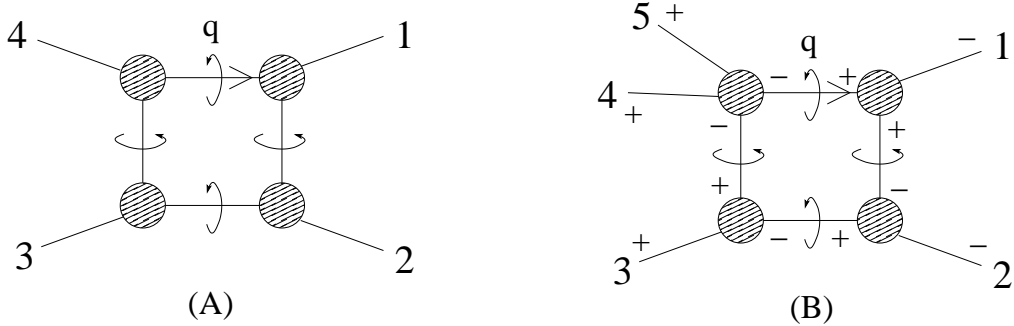}
\caption{Sum over Feynman diagrams with no triangles or bubbles. (A)
Unique four-particle configuration. (B) Unique five-particle
topology with some choice of external labels. For this particular
choice of external helicities there is a single configuration of
internal helicities.} \label{fig:Lead5}
\end{figure}

For four particles there is a single planar configuration with no
triangles or bubbles. This is depicted in figure~\ref{fig:Lead5}A.
The location of the two leading singularities is: $\ell_{a\dot a} =
\alpha \lambda^{(1)}_a\tilde\lambda^{(4)}_{\dot a}$ and $\ell_{a\dot
a} = \tilde{\alpha} \lambda^{(4)}_a\tilde\lambda^{(1)}_{\dot a}$
with $\alpha = [1,2]/[4,2]$ and $\tilde\alpha = \langle
1,2\rangle/\langle 4,2\rangle$.

It is straightforward to compute the product of the four
three-particle tree-level amplitudes on each of the solutions. In
each case we find the same answer $A^{\rm tree}st$. In order to
reproduce this with scalar integrals we start with a single box and
a coefficient to be determined.

Comparing both sides evaluated on each of the two leading
singularities we find the same condition:
\begin{equation}
B = A^{\rm tree} st.
\end{equation}
This indeed gives the correct answer. As we will see, the fact that
we got the same condition by requiring the two sides to match on
each leading singularity independently is only a peculiarity of
$n=4$ and it will not be true for any $n>4$.

\subsubsection{Five Particles}

Five particles is the first non-trivial example where the conditions
coming from the two leading singularities are different. At this
point there seem to be a puzzle since requiring agreement on each
singularity independently leads to distinct coefficients for the
same scalar box integral which is indeed a contradiction.

The resolution to this puzzle is simple but subtle. It has to do
with the fact that when momenta are complex, the integrand of loop
amplitudes is invariant under the complexification of the (double
cover) of the Lorentz group, {\it i.e.} $SL(2,\Bbb{C})\times
SL(2,\Bbb{C})$. However, the integral might be invariant under
different subgroups of $SL(2,\Bbb{C})\times SL(2,\Bbb{C})$ depending
on the choice of contour\footnote{A similar situation was
encountered in the derivation of MHV diagrams from twistor string
theory in \cite{Cachazo:2004kj}.}. The physical choice of contour
requires the loop momentum to be a real vector and hence it breaks
the group $SL(2,\Bbb{C})\times SL(2,\Bbb{C})$ down to the diagonal
$SL(2,\Bbb{C})_D$ which is a double cover of the physical Lorentz
group $SO(3,1)$. In particular, on the physical contour, scalar
integrals are parity invariant. This is what fails on the individual
$T^4$ contours.

Let us illustrate this by computing a five-particle MHV amplitude at
one-loop.

Once again there is a single topology of a planar configuration of
Feynman diagrams with only boxes. This is shown in
figure~\ref{fig:Lead5}B with a particular choice of labels. In this
case, the location of the two leading singularities is
\begin{equation}
q^{(1)}_{a\dot a} = \lambda^{(1)}_a\left( \tilde\lambda^{(1)}_{\dot
a}+\frac{\langle 2,3\rangle}{\langle
1,3\rangle}\tilde\lambda^{(2)}_{\dot a}\right), \qquad
q^{(2)}_{a\dot a} = \left(
\lambda^{(1)}_{a}+\frac{[2,3]}{[1,3]}\lambda^{(2)}_{a}\right)\tilde\lambda^{(1)}_{\dot
a}
\end{equation}

Evaluating the product of the four tree-level amplitudes is again
straightforward and gives
\begin{equation}
\left.\sum_{\rm Multiplet}\prod_{i=1}^4 A^{{\rm tree}\
(i)}\right|_{q=q^{(1)}} = A^{\rm tree}_5 s_{12}s_{23}, \qquad
\left.\sum_{\rm Multiplet}\prod_{i=1}^4 A^{{\rm tree}\
(i)}\right|_{q=q^{(2)}} = 0. \label{eq:solvi}
\end{equation}

\begin{figure}
\includegraphics[scale=0.50]{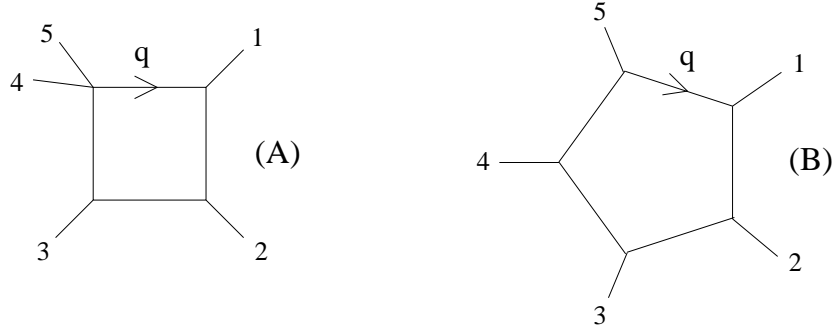}
\caption{(A) Scalar box integral for five particles. The particular
choice of labels corresponds to the integral $I^{(a):1}$ in the
text. (B) Scalar pentagon integral. This integral is denoted by
$I^{(b)}$. As in the rest of the paper, external momenta are taken
to be outgoing.} \label{fig:Lead4}
\end{figure}

We want to reproduce this behavior using scalar integrals. The
natural candidate is the scalar box integral in
figure~\ref{fig:Lead4}A. It is easy to see that on each of the
leading singularities the scalar box integral gives the same
residue, {\it i.e.}, $1/(s_{12}s_{23})$. If this box were the only
contribution on the scalar integral side we would find that by
comparing to the first equation in~(\ref{eq:solvi}) the coefficient
would be $B=A^{\rm tree}_5 s_{12}s_{23}$, which is twice the known
answer. If, instead, we use the second equation in~(\ref{eq:solvi})
we would find $B=0$ which is also wrong.

This is the puzzle mentioned earlier. In order to discover the
resolution, let us assume we did not know that ${\cal N}=4$ SYM
amplitudes can be written purely in terms of scalar
boxes\footnote{In dimensional regularization this is true only to
order $\epsilon^0$ except for $n=4$ when is true to all orders in
$\epsilon$.}. The natural starting point to reproduce the behavior
of the collection of the Feynman diagrams is the scalar box integral
in figure~\ref{fig:Lead4}. However, as we have seen, this is not
enough. Therefore we need to expand the basis. The only other scalar
integral that can contribute to the same leading singularities, but
with different residues, is a pentagon (see
figure~\ref{fig:Lead4}B).

Denoting the coefficient of the pentagon by $C$ we find that
reproducing~(\ref{eq:solvi}) means
\begin{equation}
B + \frac{C}{(q^{(1)}+k_5)^2}  =  A^{\rm tree}s_{12}s_{23}, \qquad B
+ \frac{C}{(q^{(2)}+k_5)^2}  =  0. \label{eq:juno}
\end{equation}
where $(q+k_5)^2$ is the only propagator of the pentagon not shared
by the box.

These equations can easily be solved. A convenient way to express
the solution is obtained by making the following definitions
\begin{equation}
\beta_{i} := \left(1+ \frac{\langle i+2,i+3\rangle [i+2,i] }{\langle
i+1,i+3\rangle [i+1,i]} \right)^{-1}, \qquad \tilde\beta_{i} :=
\left(1+ \frac{\langle i+2,i\rangle [i+2,i+3] }{\langle i+1,i\rangle
[i+1,i+3]} \right)^{-1}.
\end{equation}
Note that for real momenta, $\tilde\beta_i$ is the complex conjugate
of $\beta_i$. This will be important when identifying odd and even
pieces under parity.

The solution to the equations in~(\ref{eq:juno}) is
\begin{equation}
B = -A^{\rm tree}
\frac{s_{12}s_{23}\tilde\beta_5}{\beta_5-\tilde\beta_5}, \qquad C =
A^{\rm tree}\frac{s_{51}s_{12}s_{23}}{\beta_5-\tilde\beta_5}.
\label{eq:upsi}
\end{equation}

The attentive reader might anticipate another possible
contradiction: If we have fixed the coefficient of the pentagon by
using the leading singularity in figure~\ref{fig:Lead5}B, it is hard
to believe that it will simultaneously solve the equations coming
from figure~\ref{fig:Lead5}B after a cyclic permutation of the
labels\footnote{Note that the pentagon is invariant under a cyclic
permutation of the labels.}. For this to be true the same
coefficient of the pentagon should work in all cases, {\it i.e.},
$C/A^{\rm tree}$ in~(\ref{eq:upsi}) must be invariant under cyclic
permutations of the labels of the external particles!

Indeed, an explicit computation reveals that
\begin{equation}
{\cal C}:=\frac{C}{A^{\rm tree}} =
\frac{s_{i,i+1}s_{i+1,i+2}s_{i+2,i+3}}{\beta_i-\tilde\beta_i}
\label{eq:surprise}
\end{equation}
is the same quantity for all $i\in \{1,2,3,4,5\}$. How to write
${\cal C}$ in a manifestly invariant form will become clear in the
next section.

We are ready to write down the final answer for the amplitude. In
order to compare with the known result it is convenient to separate
contributions into parity even and parity odd pieces. Note that the
coefficient of the pentagon has definite parity; it is parity odd.
The coefficient of the box does not have definite parity. In order
to decompose it we write in the numerator of $B$, $\tilde\beta_5 =
(\tilde\beta_5 -\beta_5)/2+(\tilde\beta_5+\beta_5)/2$. The amplitude
is then given as
\begin{equation}
\frac{A^{(1)\rm MHV}_5}{A^{\rm tree MHV}_5} = \sum_{cyclic}\left(
\frac{1}{2}s_{12}s_{23}I^{(a):1} -
\frac{1}{2}\left(\frac{\beta_5+\tilde\beta_5}{\beta_5-\tilde\beta_5}\right)s_{12}s_{23}I^{(a):1}\right)
+ {\cal C}\; I^{(b)}. \label{eq:wewe}
\end{equation}
where $I^{(a):i}$ is a scalar box integral with three massless legs
given by the momenta of the $i^{\rm th}$, $(i+1)^{\rm th}$ and
$(i+2)^{\rm th}$ particles while $I^{(b)}$ is a pentagon with all
massless legs as shown in figure~(\ref{fig:Lead4})B.

We claim that this is the correct answer. In order to prove it we
have to go back to the usual contour of integration where $\ell$ is
a real vector and where integrals must be dimensionally regulated.
In that case we have to show that~(\ref{eq:wewe}) reduces to
\begin{equation}
\left.\frac{A^{(1)\rm MHV}_5}{A^{\rm tree MHV}_5}\right|_{\rm
Dim.Reg.} = \sum_{cyclic}\left( \frac{1}{2}s_{12}s_{23}I^{(a):1}
\right).
\end{equation}
Comparing our answer~(\ref{eq:wewe}) with the answer for real $\ell$
and in dimensional regularization we conclude that the scalar box
and pentagon integrals when dimensionally regulated must satisfy the
following identity
\begin{equation}
I^{(b)} = \frac{1}{2}\sum_{i=1}^5\left(
\frac{\beta_{i-1}+\tilde\beta_{i-1}}{s_{i-1,i}}\right)I^{(a):i}.
\label{eq:lilo}
\end{equation}
In other words, a pentagon can be written as a sum of five boxes
with particular coefficients.

This is an example of what is known as a reduction formula
\cite{Passarino:1978jh,van Neerven:1983vr,Bern:1993kr}. Indeed, one
can check that the coefficients in~(\ref{eq:lilo}) agree with those
obtained in \cite{Bern:1993kr} by using differential equations, {\it
i.e.},
\begin{equation}
\frac{\beta_{i}+\tilde\beta_{i}}{s_{i,i+1}} =
(\alpha_{i-2}-\alpha_{i-1}+\alpha_i-\alpha_{i+1}+\alpha_{i+2})\alpha_1
\alpha_2\alpha_3\alpha_4\alpha_5 s_{i+1,i+2}s_{i+2,i+3}
\end{equation}
with $\alpha_i = s_{i+1,i+2}s_{i+2,i+3}/\Xi$ and $\Xi =
\sqrt{-s_{12}s_{23}s_{34}s_{45}s_{51}}$. If we had known the
existence of reductions formulas but not their form then the leading
singularities would have given an elementary way of finding
them\footnote{In comparing with \cite{Bern:1993kr}, we had to shift
the index $i$ because what we call $I^{(a):i}$ is called
$I^{(a):i-1}$ in \cite{Bern:1993kr}.}.

Let us conclude this section by explaining why reduction formulas do
not hold on the $T^4$ contours used to compute independent leading
singularities. First observe that the choice of one such contour
breaks the $\Bbb{Z}_2$ symmetry that exchanges the two factors in
$SL(2,\Bbb{C})\times SL(2,\Bbb{C})$ which is the symmetry of the
integrand. This is nothing but a parity transformation. The reason
this affects the reduction formulas is that their derivation relies
on the identity~\cite{van Neerven:1983vr}
\begin{equation}
\int_\Gamma d^4\ell \frac{\epsilon_{\mu\nu\rho\sigma}\ell^\mu
R_1^\nu R_2^\rho
R_3^\sigma}{\ell^2(\ell+R_1)^2(\ell+R_2)^2(\ell+R_3)^2} = 0
\label{eq:redu}
\end{equation}
with $\Gamma$ a real contour\footnote{If any of the $R_i^2$ vanishes
this integral is divergent. The analysis in dimensional
regularization was performed in \cite{Bern:1993kr}. Out $T^4$
contour renders all integrals finite and this is why we do not need
any regulators.}.

A simple way to prove this is by noting that
\begin{equation}
I^\mu = \int_{\Gamma} d^4\ell
\frac{\ell^\mu}{\ell^2(\ell+R_1)^2(\ell+R_2)^2(\ell+R_3)^2} = A_1
R_1^\mu +A_2 R_2^\mu + A_3 R_3^\mu \label{eq:loren}
\end{equation}
for some scalar functions $A_i$. This is obviously true by Lorentz
invariance including parity invariance. In general any four vector,
in particular, $I^\mu$ can be expanded in a basis of vectors given
by $R_1^\mu$, $R_2^\mu$, $R_3^\mu$ and
$\epsilon^{\mu\nu\rho\sigma}R_{1\nu} R_{2\rho} R_{3\sigma}$. The
latter does not contribute in (\ref{eq:loren}) because it is not
parity invariant.

Now it is clear that on a given $T^4$ contour corresponding to a
single leading singularity the integral is not parity invariant and
(\ref{eq:loren}) does not hold due to the presence of the extra
vector $\epsilon^{\mu\nu\rho\sigma}R_{1\nu} R_{2\rho} R_{3\sigma}$
in the expansion of $I^\mu$. This also shows why in the original
quadruple cut technique introduced in \cite{Britto:2004nc}, which is
equivalent to summing over both contributions, reduction formulas
are valid. Summing over both contours preserves parity since a
parity transformation corresponds to exchanging the two $T^4$'s.

\subsubsection{Higher Point Amplitudes}

Repeating the procedure we applied to the five-particle case to
higher point amplitudes one should find reduction formulas for
higher point scalar integrals since we know that in dimensional
regularization all n-particle amplitudes are given in terms of
boxes. A case of special interest in the following is $n=6$. We
postpone the discussion to section~\ref{sec:peek} where several
coefficient of two-loop six-particles amplitudes are studied.

\section{Leading Singularity at Two Loops}

The physical meaning of leading singularities at higher loops was
given in \cite{Cachazo:2008dx}. It was shown that Feynman diagrams
arrange themselves so that a meaningful $T^{4L}\cong (S^1)^{4L}$
contour can be defined at $L$ loops. A generalization of what was
done at one-loop would require interpreting the integral over $L$
loop momenta as a contour integral in $\Bbb{C}^{4L}$. However,
combining Feynman diagrams that share the topology of a graph with
only boxes generically gives rise to less than $4L$ propagators.
Therefore defining a $T^{4L}$ contour is more subtle than in the
one-loop case. As shown in \cite{Buchbinder:2005wp}, once a partial
integration is done on one of the loop variables, new
propagator-like singularities appear which can be used to define the
$T^{4L}$ contour iteratively. The physical meaning given in
\cite{Cachazo:2008dx} is that after the partial integration, one
produces a set of Feynman diagrams at one loop order less than the
original ones and which possesses new tree-level factorization
channels. Those channels are the hidden singularities needed to
define the $T^{4L}$ contour. In this section we restrict our
attention to $L=2$. We start with the $n=4$ case in order to clarify
the concepts just explained.

\subsection{Four Particles}

\begin{figure}
\includegraphics[scale=0.50]{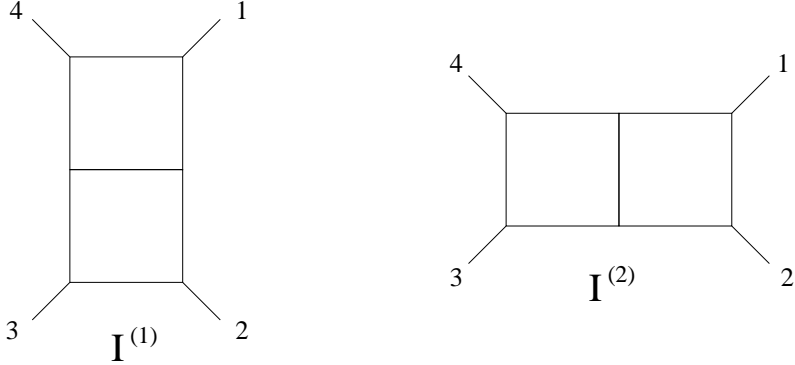}
\caption{Basis of scalar two-loop integrals for four-particle
amplitudes in ${\cal N}=4$ SYM.} \label{fig:Lead1}
\end{figure}

We will illustrate the use of the leading singularity technique on a
well known case: Four-particle two-loop amplitudes. These were first
computed in \cite{Bern:1997nh,Bern:1998ug} and are given by
\begin{equation}
A^{(2)}_4  =   A^{\rm tree}_4\left( st^2 I^{(1)} + s^2t I^{(2)}
\right) \label{eq:koki}
\end{equation}
where the integrals $I^{(1)}$ and $I^{(2)}$ are shown in figure
~\ref{fig:Lead1}.

We now proceed to reproduce this result using the leading
singularities.

In ${\cal N}=4$ SYM if we sum over Feynman diagrams and consider
contour of integrations where all legs attached to a one-loop
subdiagram are on-shell, such a subdiagram does not contain
triangles or bubbles. This means that in order to study all leading
singularities, we only need to consider sums of Feynman diagrams
that contain only boxes. In the case at hand, these are Feynman
diagrams with the topology of a 2-loop ladder.

\begin{figure}
\includegraphics[scale=0.50]{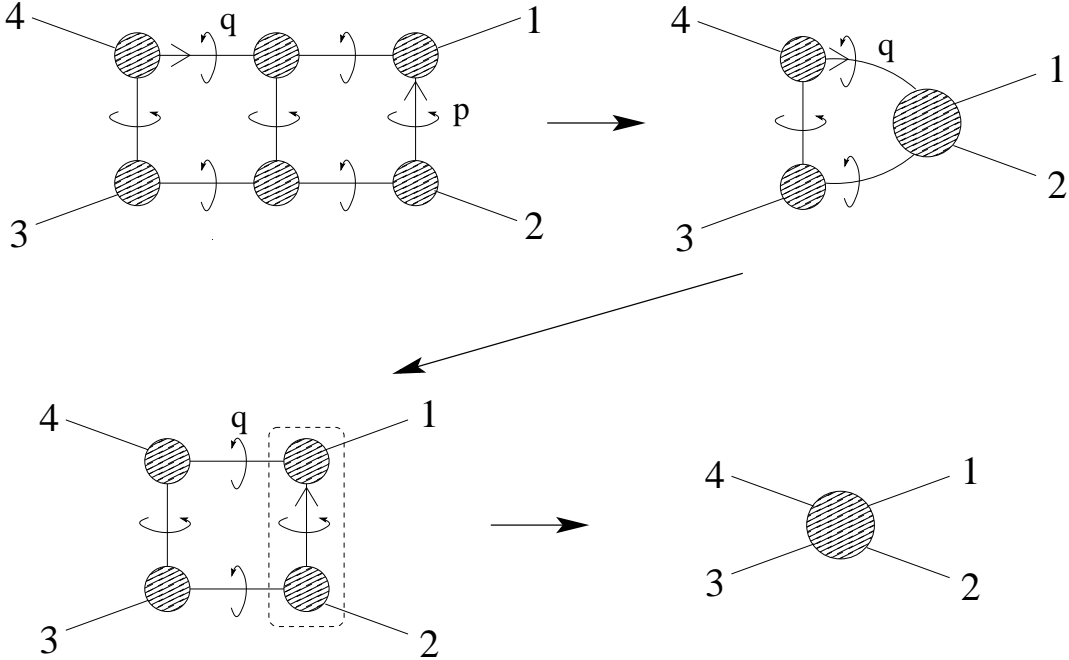}
\caption{Computation of the amplitude using the leading singularity
of Feynman diagrams. The residue of the $p$ integral on any of the
two $T^4$'s represented by the lines circling the propagators is a
tree-level four particle amplitude. Choosing a $T^4$ in the $q$
variable which induces a factorization in the $(q-k_1)^2$ channel of
the tree amplitude one finds the diagram enclosed by dashed lines.
The residue on the final $T^4$ is again a four-particle tree-level
amplitude.} \label{fig:Lead2}
\end{figure}

A particular choice of external particles is shown in
figure~\ref{fig:Lead2}. Performing the integration over the $p$
momentum we find that the product over the four three-particle
tree-level amplitudes, including the jacobian, gives rise to a
four-particle tree-level amplitude \cite{Cachazo:2008dx}. Therefore,
we are left with the sum over one-loop Feynman diagrams shown on the
upper right of figure~\ref{fig:Lead2}. Note that we have used that
for $n=4$ the product of amplitudes gives the same answer when
evaluated on the two solutions $p^{(1)}$ and $p^{(2)}$.

The tree-level four particle amplitude has two factorization
channels. One of them is in the limit when $(k_1+k_2)^2\rightarrow
0$ while the other is in the limit when $(q-k_1)^2\rightarrow 0$. It
must now be clear how to define the remaining $T^4$ in order to
perform the $q$ integration; one uses the original three propagators
together with the new propagator $1/(q-k_1)^2$. On this contour the
tree amplitude factorizes and gives rise to the diagram in the
bottom left of figure~\ref{fig:Lead2}. This is identical to the
one-loop case. Once again there are two solutions $q^{(1)}$ and
$q^{(2)}$. On both solutions the diagrams evaluate to\footnote{The
computation also involves the jacobian. In other words, we are
computing the full residue.} $A^{\rm tree}$.

On the scalar integral side, we start with an integral of the form
$I^{(2)}$ shown in figure~\ref{fig:Lead1}. After performing the
integration over $p$ one finds that the jacobian gives rise to a
factor of $1/(s(q-k_1)^2)$. This means that the scalar integral has
a non-zero residue on the same $T^8$ used for the evaluation of the
Feynman diagrams. One finds that the total integration gives
$1/(s^2t)$.

Comparing both sides we conclude that the coefficient of $I^{(2)}$
is $s^2t A^{\rm tree}_4$.

By using a cyclic permutation of the labels we find that the
coefficient of the integral $I^{(1)}$ must be $st^2 A^{\rm tree}_4$.
Combining the results we reproduce the known answer~(\ref{eq:koki}).

\section{Two-Loop Five-Particle Amplitude}

Five-particle MHV two-loop amplitudes in ${\cal N}=4$ SYM were
computed in \cite{Bern:2006vw} using the unitarity based method.
Here we write the known answer and then show how to re-derive it
using the leading singularity technique. The answer we find comes
out in an strikingly different form. It is important to make a
distinction between MHV and ${\overline{\rm MHV}}$ as the two
amplitudes, when normalized by $A^{\rm tree}_5$, are different.

The expression given in \cite{Bern:2006vw} is
\begin{equation}
\begin{array}{ccl}
A^{(2){\rm MHV}}_5 & = &  \frac{1}{8} A^{\rm tree, MHV}_5\sum_{\rm
cyclic}\left(
s_{12}^2s_{23}I^{(a)}(\epsilon)+s_{12}^2s_{15}I^{(b)}(\epsilon)+s_{12}s_{34}s_{45}I^{(c)}(\epsilon)
\right.\\
&  & \left. + R\left[ 2I^{(d)}(\epsilon) -2s_{12}I^{(e)}(\epsilon) +
\frac{s_{12}}{s_{34}s_{45}}\left(
\frac{\delta_{-++}}{s_{23}}I^{(b)}(\epsilon)-\frac{\delta_{-+-}}{s_{51}}I^{(a)}(\epsilon)\right)
+ \frac{\delta_{+-+}}{s_{23}s_{51}}I^{(c)}(\epsilon) \right]\right)
\end{array}
\label{eq:theirs}
\end{equation}
where
\begin{equation}
\begin{array}{ccl}
R & = & \epsilon_{1234}s_{12}s_{23}s_{34}s_{45}s_{51}/G_{1234},\\
\delta_{abc} & = &
s_{51}s_{12}+as_{12}s_{23}+bs_{23}s_{34}-s_{45}s_{51}+cs_{34}s_{45},\\
\epsilon_{1234} & = & 4i\varepsilon_{\mu\nu\rho\sigma}k_1^\mu
k_2^\nu k_3^\rho k_4^\sigma = {\rm tr}\left[\gamma_5
\sk_1\sk_2\sk_3\sk_4 \right],
\end{array}
\label{eq:fivebrane}
\end{equation}
and
\begin{equation}
G_{1234} = {\rm det}\left( \begin{array}{cccc} 0 & s_{12} & s_{13} & s_{14}\\
s_{12} & 0 & s_{23} & s_{34} \\ s_{13} & s_{23} & 0 & s_{34} \\
s_{14} & s_{24} & s_{34} & 0
\end{array}\right)
\end{equation}

\begin{figure}
\includegraphics[scale=0.50]{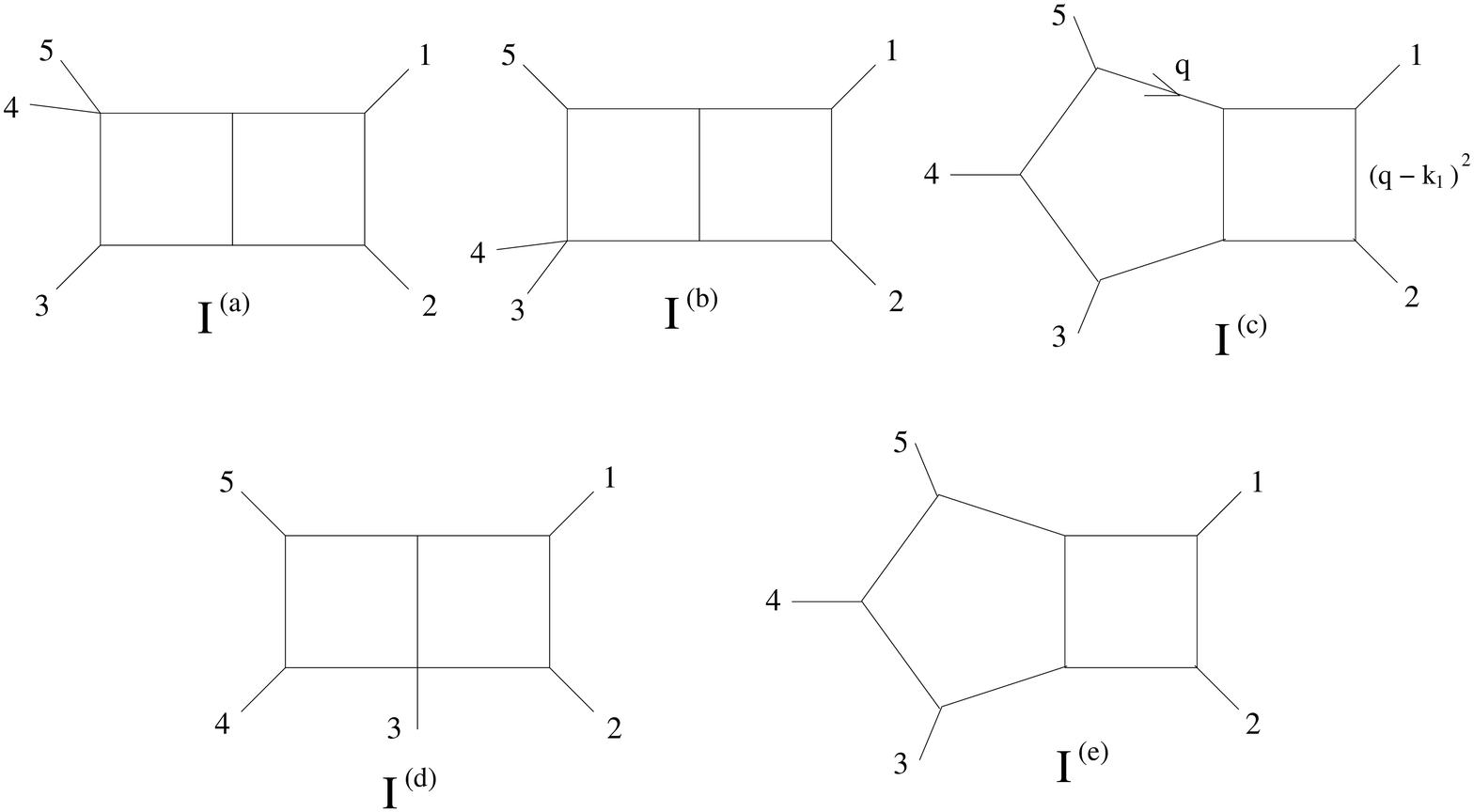}
\caption{Basis of two-loop integrals for five-particle amplitudes in
${\cal N}=4$ SYM. Only one choice of external labels is shown. The
full basis is obtained by considering all cyclic permutation of
labels.} \label{fig:Lead3}
\end{figure}

All integrals $I^{(a)}$, $I^{(b)}$, $I^{(c)}$, $I^{(d)}$ and
$I^{(e)}$ are defined for a particular choice of external particles
as shown in figure~\ref{fig:Lead3}.

\subsection{Computation Using The Leading Singularity Technique}

We start by following the same steps as in the four-particle case.
There are three different topologies of diagrams with only boxes. We
have to analyze all of them and built an expression in terms of
scalar integrals which reproduces the behavior of Feynman diagrams
when evaluated in all possible leading singularities.

\begin{figure}
\includegraphics[scale=0.50]{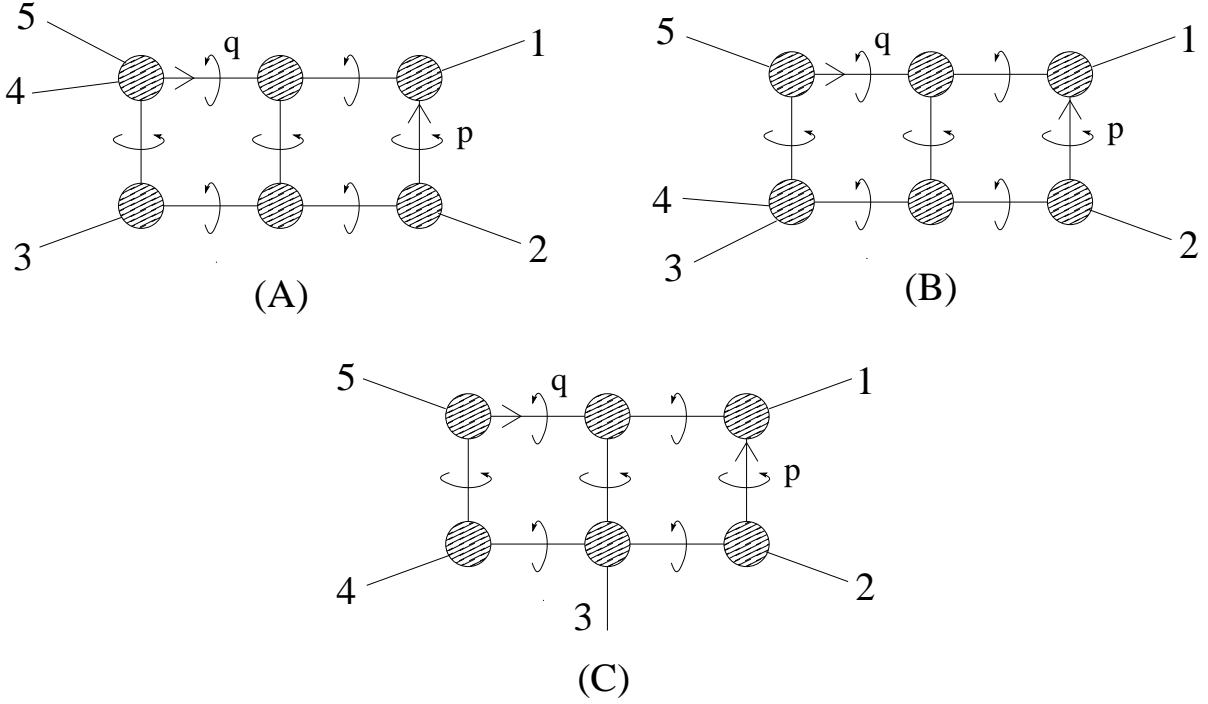}
\caption{All inequivalent topologies of sums of Feynman diagrams
with only boxes for five particles at two loops. A particular choice
of external labels is shown. All cyclic permutations of the labels
must also be considered.} \label{fig:Lead7}
\end{figure}

The three topologies are depicted in figure~\ref{fig:Lead7} where a
particular choice of external labels was made. The total set of
configurations is then obtained by cyclic permutations of the
labels.

\subsubsection{First Topology}

Consider the set of Feynman diagrams in figure~\ref{fig:Lead7}A.
This is the first of the three different topologies that contain
only boxes for five particles.

\begin{figure}
\includegraphics[scale=0.50]{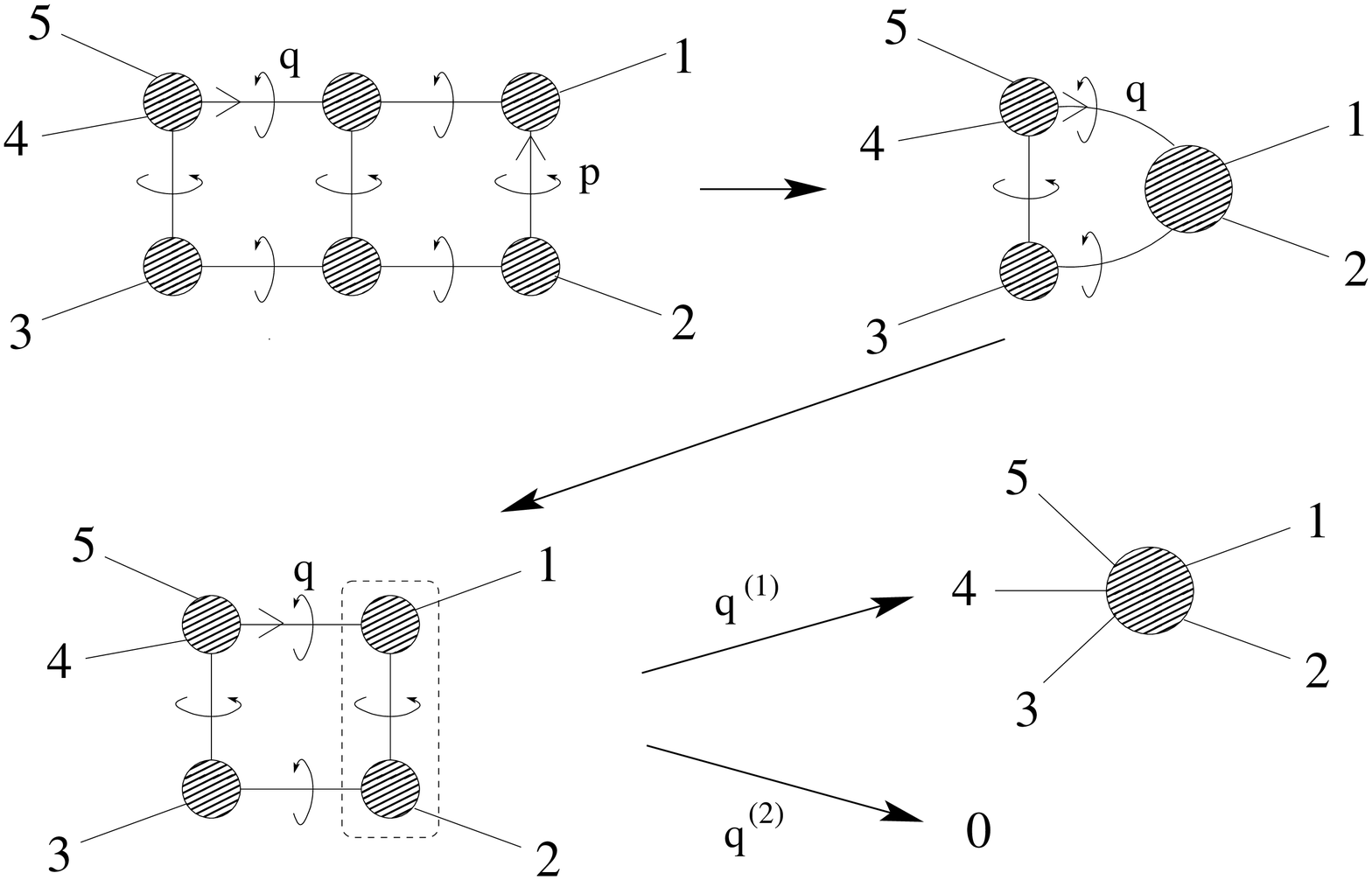}
\caption{Evaluation of the sum over Feynman diagrams on two
different $T^8$ contours. The residue of the $p$ integral on any of
the two $T^4$'s represented by the lines circling the propagators is
a tree-level four particle amplitude. For $n=5$, as opposed to
$n=4$, there are two choices for the $T^4$ contour in the $q$
variable. Here we choose the one that induces a factorization in the
$(q-k_1)^2$ channel of the tree amplitude. The other $T^4$ is also
important and is considered in figure \ref{fig:Lead9}. The residue
on the final $T^4$ is the five-particle tree-level amplitude on
$q^{(1)}$ and zero on $q^{(2)}$.} \label{fig:Lead6}
\end{figure}

Carrying out the integration over a $T^4$ contour in the $p$
variables we find that for any of the two solutions we get a
four-particle tree-level amplitude with $1$ and $2$ as external legs
(see figure~\ref{fig:Lead6}). In order to continue, we choose a
$T^4$ contour in the $q$ variables that induces a factorization of
the four-particle amplitude that contains the external particles $1$
and $2$. Note that in this case there is a second four-particle
amplitude; the one that contains external particles $4$ and $5$. A
$T^4$ which induces a factorization of the second four-particle
amplitude must also be considered and we postpone its analysis to
section~\ref{sec:adi}. Note that this second possibility was not
present in the four-particle case.

From the bottom left of figure~\ref{fig:Lead6} we find that the
problem at hand is identical to the one-loop five-particle amplitude
discussed in the previous section. The difference between the two
cases comes in analyzing the scalar integrals. By analogy with the
one loop case, we start by assuming that the basis contains
integrals $I^{(a)}$ and $I^{(e)}$ in figure~\ref{fig:Lead3}. After
carrying out the integration over the first $T^4$ in the scalar
integrals, we find that the problem also reduces to that of the
one-loop case except that the coefficients are multiplied by an
extra factor of $1/s_{12}$. In other words, if we denote the
coefficients of the two-loop integrals $I^{(a)}$ and $I^{(e)}$ by
$B^{\rm 2-loop}$ and $C^{\rm 2-loop}$ respectively, then the
equations are
\begin{equation}
\frac{B^{\rm 2\h loop}}{s_{12}} + \frac{C^{\rm 2\h
loop}}{s_{12}(q^{(1)}+k_5)^2} = A^{\rm tree}s_{12}s_{23}, \qquad
\frac{B^{\rm 2\h loop}}{s_{12}} + \frac{C^{\rm 2\h
loop}}{s_{12}(q^{(2)}+k_5)^2} = 0. \label{eq:popi}
\end{equation}
The solution is obtained by making the substitution $B \rightarrow
B^{\rm 2\h loop}/s_{12}$ and $C \rightarrow C^{\rm 2\h loop}/s_{12}$
in~(\ref{eq:upsi}), {\it i.e,}
\begin{equation}
B^{\rm 2\h loop} = - A^{\rm tree}
\frac{s_{12}^2s_{23}\tilde\beta_5}{\beta_5-\tilde\beta_5}, \qquad
C^{\rm 2\h loop} = A^{\rm
tree}\frac{s_{51}s_{12}^2s_{23}}{\beta_5-\tilde\beta_5}.
\label{eq:lala}
\end{equation}

Recall that the coefficient of the pentagon $C$ at one loop was
invariant under cyclic permutations of the labels. Here the symmetry
has been explicitly broken by the extra factor of $s_{12}$ and hence
the integral $I^{(e)}$ must be inside the sum over cyclic
permutations.

\subsubsection{Second Topology}

The next topology of Feynman diagrams is shown in
figure~\ref{fig:Lead7}B. The contributing integrals are clearly
$I^{(b)}$ and $I^{(e)}$ in figure~\ref{fig:Lead3}. After performing
the integration over $p$ on a $T^4$ contour and choosing the
remaining $T^4$ in the $q$ variables to induce a factorization on
the four-particle amplitude with $1$ and $2$, we go back to a
one-loop calculation related to the one done in the previous case by
a cyclic permutation.

The solution for the coefficients is obtained by performing a cyclic
permutation on the one-loop coefficients and {\it then} multiplying
by the $1/s_{12}$ factor to convert them into two-loop coefficients.
The explicit form of the coefficients is
\begin{equation}
B^{\rm 2\h loop} = - A^{\rm tree}
\frac{s_{51}s_{12}^2\tilde\beta_4}{\beta_4-\tilde\beta_4}, \qquad
C^{\rm 2\h loop} = A^{\rm
tree}\frac{s_{45}s_{51}s_{12}^2}{\beta_4-\tilde\beta_4}.
\label{eq:polis}
\end{equation}

Note that we have computed the coefficient of the pentagon-box
integral, $I^{(e)}$, once again and the answer looks very different.
Using the same identity~(\ref{eq:surprise}) as in the previous
section it is clear that the two formulas are the same since all we
have done is to multiply them by the same $s_{12}$ factor.

\subsubsection{Third Topology}

\begin{figure}
\includegraphics[scale=0.45]{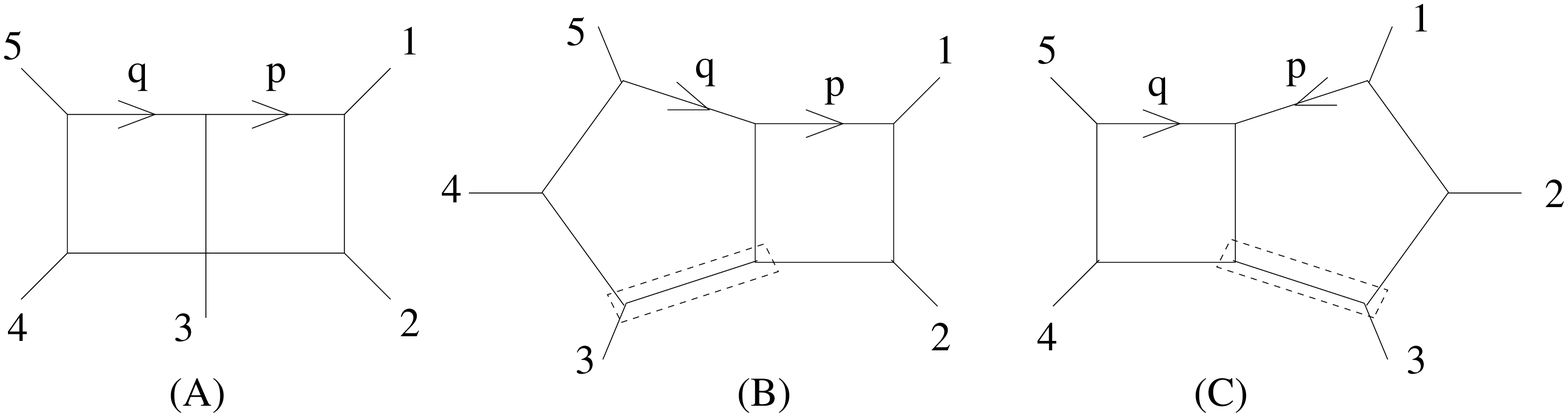}
\caption{Scalar integrals contributing to the third kind of
topology. In the computation of the residues the propagators not
used as poles remain and must be evaluated at the location of the
singularity. There is one propagator left in each pentagon-box
integral and they are enclosed by dashed lines in the figure.}
\label{fig:Lead8}
\end{figure}

Let us consider the final topology shown in figure~\ref{fig:Lead7}C.
The possible scalar diagrams that contribute to the $T^8$ which
computes the natural leading singularity of this topology are shown
in~\ref{fig:Lead8}. Let us denote their coefficients by $B$, $C_5$
and $C_3$ respectively. The coefficients $C_3$ and $C_5$ have
already been computed. So in practice one would need a single
equation to determine $B$. Let us however consider the two equations
that arise and check that they can be consistently solved.

The situation here is different from the situation in the previous
cases since the sum over Feynman diagrams has zero residue on the
leading singularities we will consider. From that point of view, we
can think of the integral in figure~\ref{fig:Lead8}A as canceling
unphysical singularities in the integrals in
figures~\ref{fig:Lead8}B and ~\ref{fig:Lead8}C which were already
shown to be present in the amplitude.

The location of the unphysical leading singularities is given by
\begin{equation}
p^{(1)} = \frac{[1,2]}{[5,2]}\lambda_1\tilde\lambda_5,\quad q^{(1)}
= -\frac{\langle 5,4\rangle}{\langle
1,4\rangle}\lambda_1\tilde\lambda_5, \label{eq:oli}
\end{equation}
and
\begin{equation}
p^{(2)} = \frac{\langle 1,2\rangle}{\langle
5,2\rangle}\lambda_5\tilde\lambda_1,\quad q^{(2)} = -\frac{[ 5,4]}{[
1,4]}\lambda_5\tilde\lambda_1. \label{eq:symi}
\end{equation}

In order to understand how these arise note that the middle
propagator, $1/(q-p)^2$, in figure ~\ref{fig:Lead8}A has two poles
on the locus where $p^2=q^2=0$. These are $\langle p,q\rangle =0$
and $[p,q]=0$. Using both of them at the same time gives the $T^8$
used to obtain (\ref{eq:oli}) and (\ref{eq:symi}).

Since $p$ and $q$ are proportional to each other in both solutions
it is clear that the sum over Feynman diagrams shown in
figure~\ref{fig:Lead7}C vanishes.

The equations we have to solve take the form
\begin{equation}
B + \frac{C_5}{(q^{(i)}-k_1-k_2)^2} +
\frac{C_3}{(p^{(i)}-k_2-k_3)^2} = 0
\end{equation}
for $i=1,2$. The consistency condition that the known coefficients
$C_3$ and $C_5$ must satisfy for these equations to have a solution
is
\begin{equation}
\frac{C_5}{(q^{(1)}-k_1-k_2)^2} + \frac{C_3}{(p^{(1)}-k_2-k_3)^2}
=\frac{C_5}{(q^{(2)}-k_1-k_2)^2} + \frac{C_3}{(p^{(2)}-k_2-k_3)^2}.
\label{eq:colli}
\end{equation}
Using the explicit form of the coefficients
\begin{equation}
C_3 = A^{\rm
tree}\frac{s_{34}s_{45}^2s_{51}}{\beta_3-\tilde\beta_3}, \qquad C_5
=A^{\rm tree}\frac{s_{51}s_{12}^2s_{23}}{\beta_5-\tilde\beta_5}
\end{equation}
one can check that~(\ref{eq:colli}) is indeed satisfied. Therefore
any of the two equations determine $B$. Choosing $i=1$ gives
\begin{equation}
B = - \frac{C_5}{(q^{(1)}-k_1-k_2)^2} -
\frac{C_3}{(p^{(1)}-k_2-k_3)^2}.
\end{equation}
This expression can be dramatically simplified if~(\ref{eq:colli})
is used to solve for $C_3$ in terms of $C_5$. The answer turns out
to be
\begin{equation}
B=-\frac{C_5}{s_{12}}.
\end{equation}
This is the coefficient of integral $I^{(d)}$ in figure
\ref{fig:Lead3}.

\subsubsection{Additional Leading Singularities}
\label{sec:adi}

As mentioned in the discussion of the first kind of topology, once
the integration over the $T^4$ corresponding to the $p$ variables is
carried out, one is left with a sum over Feynman diagrams that
possesses two four-particle amplitudes (see figure~\ref{fig:Lead6});
one which contains particles $1$ and $2$ and another which contains
particles $4$ and $5$. Previously, we chose the remaining $T^4$
contour such that the amplitude with $1$ and $2$ factorizes. Now we
have to check that the leading singularity corresponding to $4$ and
$5$ factorizing also works.

Here we get four equations corresponding to any combination of
$q^{(1)}$, $q^{(2)}$ and $p^{(1)}$, $p^{(2)}$. Note that none of the
scalar integrals with the topology of two boxes, {\it i.e.}
$I^{(a)}$, $I^{(b)}$, $I^{(d)}$ contributes to these leading
singularities. At this point, we have a single integral that
contributes, {\it i.e.}, $I^{(e)}$. Not surprisingly, it is not
possible to solve all the equations with the coefficient of a single
integral. This means that at least one more integral is missing.

The requirements for the new integral are:

\begin{itemize}

\item It must be non-zero on all the new $T^8$ contours.

\item It must vanish on all the previous $T^8$ contours since they have
already been accounted for.

\end{itemize}

The only possibility is to start with an integral of the form
$I^{(e)}$ which ensures the first requirement and then introduce a
zero in the numerator with removes the pole $(q-k_1)^2$ which enters
in the definition of all other $T^8$'s already studied. Removing the
pole at $(q-k_1)^2$ guarantees that the new integral has zero
residue in all previous leading singularities thus fulfilling the
second requirement.

The new integral has the topology of $I^{(e)}$ and a numerator
factor $(q-k_1)^2$. This integral is precisely $I^{(c)}$ in
figure~\ref{fig:Lead3}.

Let us show that by adding $I^{(c)}$ to the basis all remaining
leading singularities can be accounted for. Let its coefficient be
$D$.

The four leading singularities are located at
\begin{equation}
p^{(1)}(q) =\frac{[1,2]}{[q,2]}\lambda_1\tilde\lambda_q, \quad
p^{(2)}(q) =\frac{\langle1,2\rangle}{\langle
q,2\rangle}\lambda_q\tilde\lambda_1,
\end{equation}
and
\begin{equation}
q^{(1)} = -\lambda_5\left(\tilde\lambda_5+\frac{\langle
3,4\rangle}{\langle 3,5\rangle}\tilde\lambda_4\right), \qquad
q^{(2)} = -\left(\lambda_5+\frac{[3,4]}{[
3,5]}\lambda_4\right)\tilde\lambda_5.
\end{equation}

\begin{figure}
\includegraphics[scale=0.50]{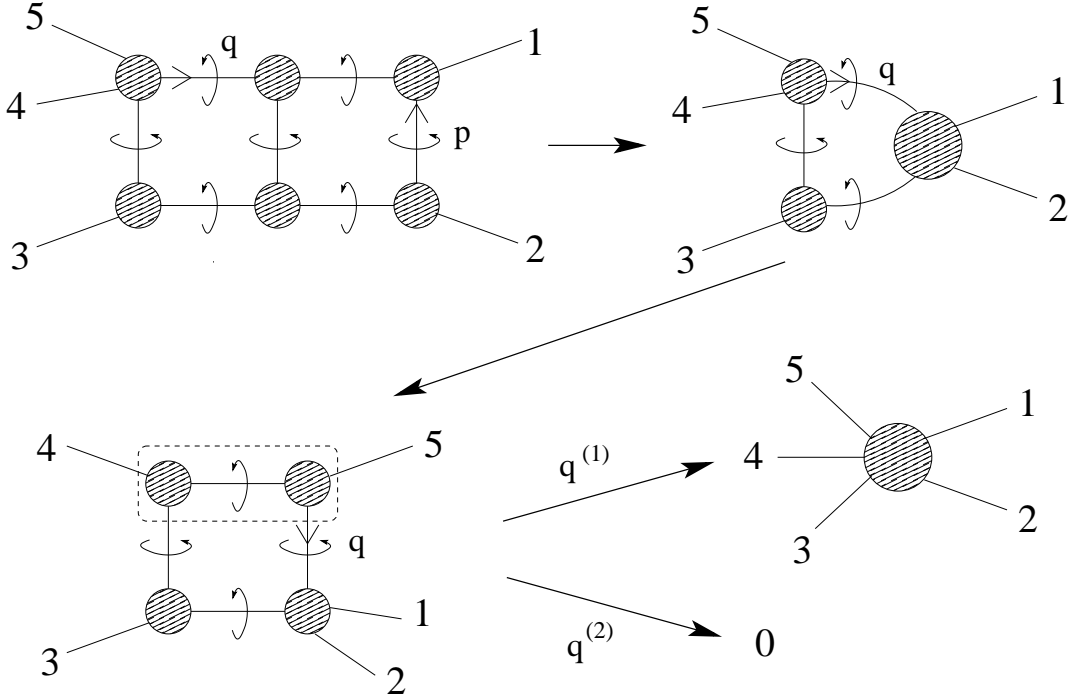}
\caption{Evaluation of the sum over Feynman diagrams over the new
$T^8$ contours. In figure \ref{fig:Lead6} we chose the second $T^4$
to induce a factorization in the $(q-k_1)^2$ channel, here we choose
it to induce a factorization of the second four-particle amplitude
in the $(q+k_5)^2$ channel.} \label{fig:Lead9}
\end{figure}

The value of $p$ does not enter in the computation and we only have
to consider the two equations coming from the two values of $q$'s.
The equations are obtained in a completely analogous way from the
previous calculations following the steps in figure~\ref{fig:Lead9}.
The two equations are
\begin{equation}
\frac{C}{s_{12}s_{34}s_{45}(q^{(1)}-k_1)^2}
+\frac{D}{s_{34}s_{45}s_{12}}= A^{\rm tree}, \qquad
\frac{C}{(q^{(2)}-k_1)^2} +D= 0.
\end{equation}
The coefficient $C$ has already been computed so in practice one
would simply use the second equation to determine $D$. Let us not do
that and solve the equations for $D$ and $C$.

Defining
\begin{equation}
\gamma := \left(1+\frac{\langle 3,4\rangle [4,1]}{\langle 3,5\rangle
[5,1]} \right)^{-1}, \qquad \tilde\gamma := \left(1+\frac{\langle
4,1\rangle [3,4]}{\langle 5,1\rangle [3,5]} \right)^{-1}
\end{equation}
the solution is given as follows
\begin{equation}
C = A^{\rm
tree}\frac{s_{34}s_{45}s_{51}s_{12}}{(\gamma-\tilde\gamma)}, \qquad
D = -A^{\rm tree}s_{12}s_{34}s_{45}\frac{\tilde\gamma}{\gamma -
\tilde\gamma}.
\end{equation}
An explicit computation reveals that the new expression for $C$
agrees with that of $C^{2\h \rm loop}$ found earlier.

\subsubsection{Final Result}

We now collect all the results obtained by requiring that all
leading singularities are reproduced correctly. Using the scalar
integrals defined in figure~\ref{fig:Lead3} the two-loop amplitude
is
\begin{equation}
\frac{A^{(2)}_5}{A^{(0)}_5} =\!\! \sum_{cyclic}\! s_{12}\left( \!
-\frac{s_{12}s_{23}\tilde\beta_5}{\beta_5-\tilde\beta_5}I^{(a)}
-\frac{s_{51}s_{12}\tilde\beta_4}{\beta_4-\tilde\beta_4}I^{(b)}-
\frac{s_{34}s_{45}\tilde\gamma}{\gamma-\tilde\gamma}I^{(c)}-
\frac{s_{51}s_{23}}{\beta_5-\tilde\beta_5}I^{(d)}
+\frac{s_{51}s_{12}s_{23}}{\beta_5-\tilde\beta_5}I^{(e)}\! \right)
\label{eq:ours}
\end{equation}
where
\begin{equation}
\beta_{i} := \left(1+ \frac{\langle i+2,i+3\rangle [i+2,i] }{\langle
i+1,i+3\rangle [i+1,i]} \right)^{-1}, \qquad  \gamma :=
\left(1+\frac{\langle 3,4\rangle [4,1]}{\langle 3,5\rangle [5,1]}
\right)^{-1},
\end{equation}
and $\tilde\beta_i$ and $\tilde\gamma$ the parity conjugated
expressions of $\beta_i$ and $\gamma$ respectively.

An even simpler expression can be obtained if one uses the different
identities found in the previous subsections to write
\begin{equation}
\frac{1}{\gamma -\tilde\gamma} =
\frac{s_{12}s_{23}}{s_{34}s_{45}}\frac{1}{(\beta_5-\tilde\beta_5)},
\qquad \frac{1}{\beta_4 -\tilde\beta_4} =
\frac{s_{23}}{s_{45}}\frac{1}{(\beta_5-\tilde\beta_5)}.
\label{eq:simp}
\end{equation}
Using (\ref{eq:simp}) in (\ref{eq:ours}) one finds
\begin{equation}
\frac{A^{(2)}_5}{A^{(0)}_5} =\! \sum_{cyclic}\;
\frac{s_{12}s_{23}}{\beta_5-\tilde\beta_5}\left( \!
-s_{12}\tilde\beta_5I^{(a)}
-\frac{s_{51}s_{12}\tilde\beta_4}{s_{45}}I^{(b)}- s_{12}\tilde\gamma
I^{(c)}- s_{51}I^{(d)} +s_{51}s_{12}I^{(e)}\! \right).
\end{equation}
In order to compare with the known answer it is convenient to split
the coefficients in parity even and odd terms. Note that using out
technique both pieces are computed simultaneously and equally
straightforwardly. The separation into definite parity terms is
easily done by recalling that for real momenta, $\tilde\beta$,
$\tilde\gamma$ are the complex conjugate to $\beta$ and $\gamma$
respectively. This means that we can write, for example,
\begin{equation}
-\frac{s_{12}^2s_{23}\tilde\beta_5}{\beta_5-\tilde\beta_5}I^{(a)} =
\frac{1}{2}s_{12}^2s_{23}\left( 1-
\frac{\beta_5+\tilde\beta_5}{\beta_5-\tilde\beta_5}\right)I^{(a)}.
\end{equation}
Note that some coefficients are naturally parity odd. Like those of
$I^{(e)}$ and $I^{(d)}$.

It is easy to check analytically using a symbolic manipulation
program like {\tt Mathematica} that, up to an overall normalization
of $1/4$, (\ref{eq:ours}) is exactly equal to~(\ref{eq:theirs}).

\section{A Peek At Two-Loop Six-Particle Amplitudes: MHV and Next-to-MHV}
\label{sec:peek}

One of the advantages of our technique is that the homogenous part
of the linear equations that determine the coefficients of the
integrals is helicity independent! All the helicity information
enters in the inhomogeneous part.

In order to illustrate this feature the first non-trivial case is
that of six particles. This is the first case where next-to-MHV
configurations are possible. In this section we choose a particular
subset of Feynman diagrams contributing to two-loop six-particle
amplitudes. Studying the leading singularities one can write down
linear equations that determine the {\it complete} coefficient of
all pentagon-pentagon integrals and of a certain class of
pentagon-box integrals. The determination of the complete set of
linear equations which gives the full amplitude is outside the scope
of this paper and we leave it for future work.

\begin{figure}
\includegraphics[scale=0.40]{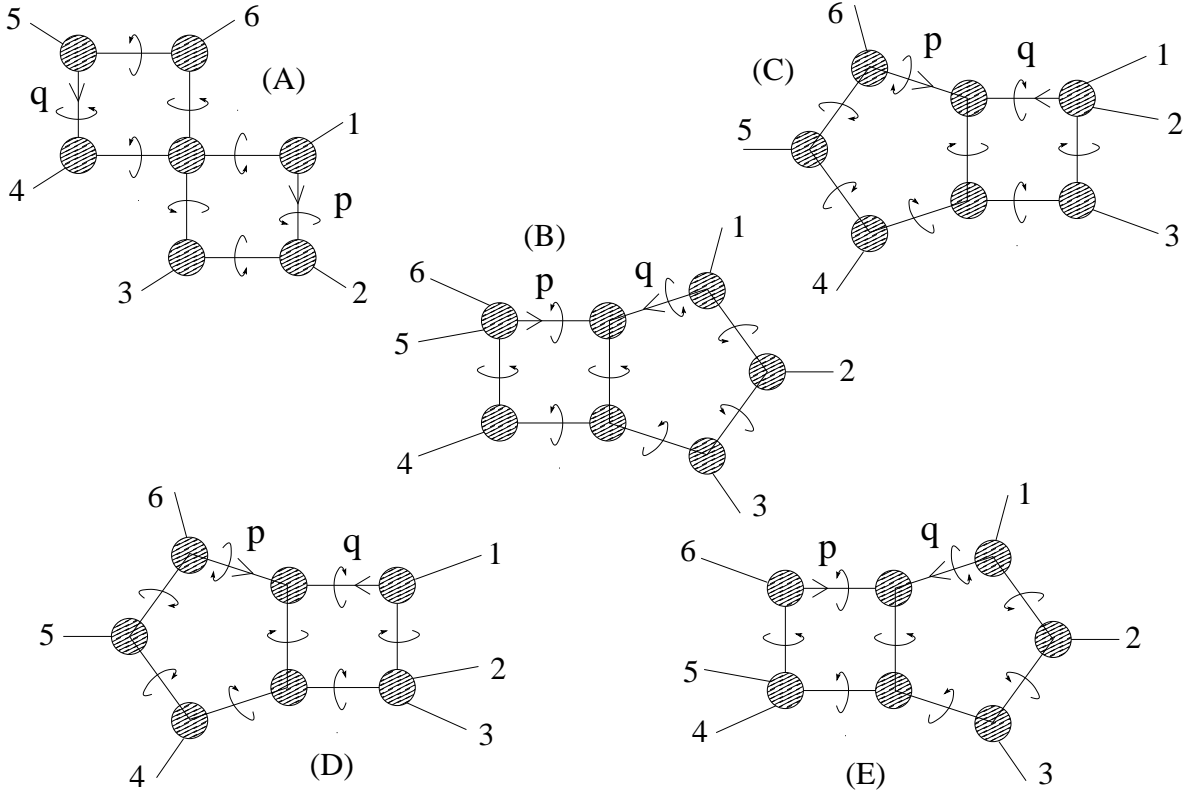}
\caption{Topologies of sums over Feynman diagrams which can be used
to determine the coefficients of all pentagon-pentagon and some
class of pentagon-box integrals in a six-particle two-loop
amplitude. Some choice of external labels has been made. No
helicities have been assigned since all choices, MHV and
next-to-MHV, can be treated simultaneously.} \label{fig:LeadSix3}
\end{figure}

The set of Feynman diagrams we consider generates the five
topologies shown in figure \ref{fig:LeadSix3}. We write down
explicitly the equations coming from $(A)$ and $(B)$, as the
remaining three, $(C)$, $(D)$, and $(E)$, are completely analogous
to $(B)$.

\subsection{Topology $(A)$}

\begin{figure}
\includegraphics[scale=0.35]{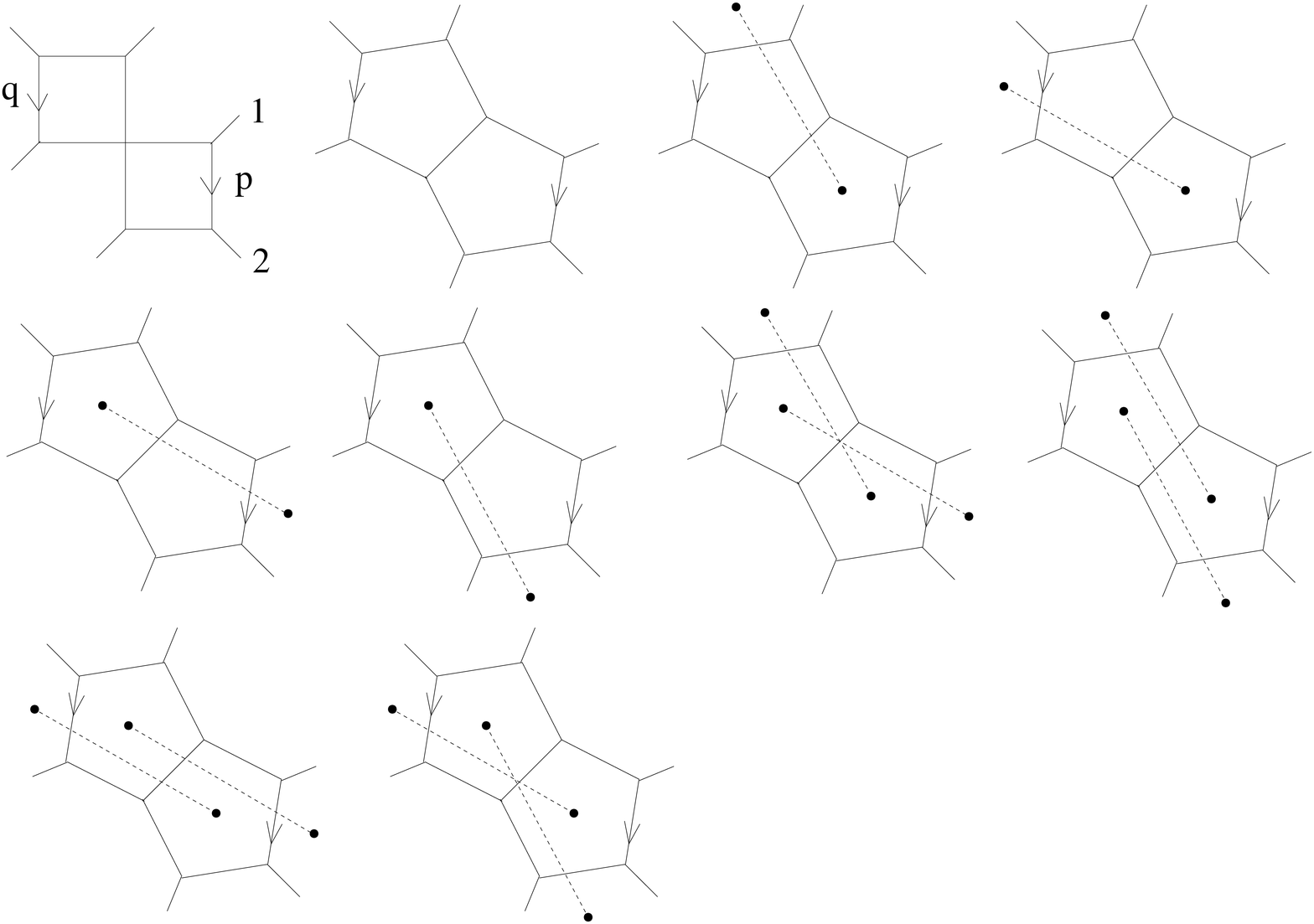}
\caption{Scalar and generalized scalar integrals that can contribute
to the first kind of topology. Their coefficients are
$B,C,E_1,E_2,E_3,E_4,D_1,D_2,D_3,D_4$ respectively. In order to
avoid cluttering of the figure we have added the minimal amount of
information needed to determine the labels. In the double box
diagram, other external legs are labeled following the color
ordering. All pentagon-pentagon diagram naturally inherit their
labeling from the double-box diagram.} \label{fig:LeadSix2}
\end{figure}

The integrals contributing to the first kind of topology are shown
in figure \ref{fig:LeadSix2}. If we were to follow the same steps as
in the previous section we would start by taking only the first two
integrals in the figure and then realize that it is not possible to
solve the four equations that arise by comparing all leading
singularities. From the experience with the five-particle case, we
start by adding eight integrals with numerators such that they will
contribute to the four leading singularities under consideration but
they will not contribute to other singularities where they are not
needed just like in the case of $I^{(c)}$ for five particles.

Using the labels in the figure, the four leading singularities are
found by choosing any combination of $p_*$'s and $q_*$'s from
\begin{equation}
p^{(1)} = \frac{\langle 2,3\rangle}{\langle
1,3\rangle}\lambda_1\tilde\lambda_2,\quad p^{(2)} =
\frac{[2,3]}{[1,3]}\lambda_2\tilde\lambda_1,\quad q^{(1)} =
-\frac{\langle 5,6\rangle}{\langle
4,6\rangle}\lambda_4\tilde\lambda_5,\quad q^{(2)} =-\frac{[ 5,6 ]}{[
4,6]}\lambda_5\tilde\lambda_4.
\end{equation}

Let us denote the evaluation of the sum over Feynman diagrams (see
figure \ref{fig:LeadSix3}A) in a particular solution $(p, q)$ by
$F_{p,q}$. Then the equations are
\begin{equation}
\begin{array}{ccl}
s_{12}s_{23}s_{45}s_{56}\; F_{p,q} & = & \left[ B +
\frac{1}{(p+q-k_{234})^2}\left(
C+(E_1+D_1(q-k_{234})^2)(p+k_{61})^2+ \right. \right.
\\ & & \left.\left. (E_3+D_2(p-k_{234})^2)(q-k_{234})^2+
(E_4+D_3(p+k_{61})^2)(q-k_{34})^2+ \right. \right.\\ & &
\left.\left. (E_2+D_4(q-k_{34})^2)(p-k_{234})^2\right)\right]
\end{array}
\label{eq:first}
\end{equation}
The labeling of the coefficients is explained in the caption of
figure \ref{fig:LeadSix2}.

The system at hand involves four equations and ten unknown
coefficients. In order to find a set of equations sufficient to
completely determine the coefficients we have to consider the other
topologies in figure \ref{fig:LeadSix3}.

Before going to the next topology, lead us compute $F(p,q)$ in some
cases in order to illustrate the procedure which turns out to be
fairly simple since the computation is reduced to that at one-loop.

\begin{figure}
\includegraphics[scale=0.40]{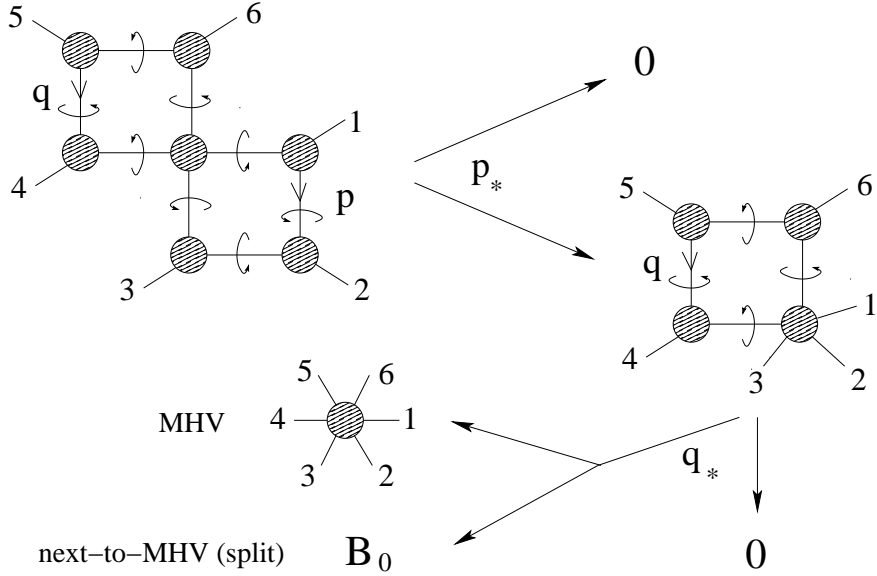}
\caption{Computation of functions $F_{p,q}$ in two cases: MHV with
helicity configuration $\{1^-,2^-,3^+,4^+,5^+,6^+\}$ and next-to-MHV
(split), {\it i.e.} $\{1^-,2^-,3^-,4^+,5^+,6^+\}$
 } \label{fig:LeadSix4}
\end{figure}

All the steps are shown in figure \ref{fig:LeadSix4}. Consider the
lower box in the two-loop diagram. These are the Feynman diagrams in
a one-loop five-particle amplitude where two of the external momenta
are actually internal legs in the two-loop diagram. From the
one-loop discussion in the previous section we know that depending
on the helicities, one solution, $p_*$, gives zero while the other
gives $A^{\rm tree}_5$. Plugging this into the two-loop diagram one
gets, in the non-zero case, a one-loop six particle diagram. If the
amplitude we are trying to compute is MHV or $\overline{\rm MHV}$
then the answer is either zero or $A^{\rm tree}_6$.

As an explicit example take helicities to be
$\{1^-,2^-,3^+,4^+,5^+,6^+\}$. Then we find
\begin{equation}
F(p^{(1)},q^{(1)}) = A^{\rm tree}_6,\quad F(p^{(1)},q^{(2)})
=F(p^{(2)},q^{(1)}) =F(p^{(2)},q^{(2)}) =0.
\end{equation}

If the amplitude is next-to-MHV then the answer is more interesting
as it corresponds to the coefficient of a particular one-mass
integral in a six-particle one-loop amplitude. All six-particle
one-loop next-to-MHV amplitudes amplitudes where computed in the
early 90's in \cite{Bern:1994cg}. They are all given in terms of a
quantity defined in eq. 6.13 of \cite{Bern:1994cg} (we have parity
conjugated the expression in \cite{Bern:1994cg})
\begin{equation}
B_0 := \frac{\langle 1|2+3|4]\langle 3|1+2|6]
t^3_{123}}{[1,2][2,3]\langle 4,5\rangle\langle
5,6\rangle(t_{123}t_{345}-s_{12}s_{45})(t_{123}t_{234}-s_{23}s_{56})}.
\end{equation}

As a explicit example take helicities to be
$\{1^-,2^-,3^-,4^+,5^+,6^+\}$. Then we find
\begin{equation}
F(p^{(1)},q^{(1)}) =F(p^{(2)},q^{(1)}) =F(p^{(2)},q^{(2)}) =0, \quad
F(p^{(1)},q^{(2)}) = B_0.
\end{equation}

It should be clear that any other helicity configurations can be
treated analogously.

\subsection{Topology $(B)$}

\begin{figure}
\includegraphics[scale=0.30]{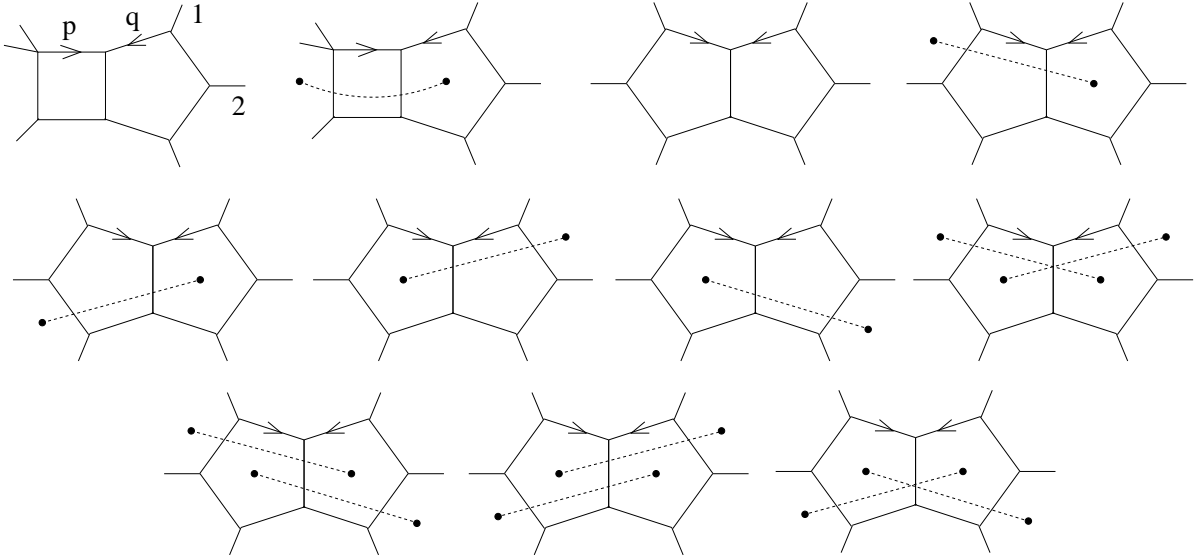}
\caption{Scalar and generalized scalar integrals that can contribute
to the second kind of topology. Their corresponding coefficients are
$F,G,C,E_1,E_2,E_3,E_4,D_1,D_2,D_3,D_4$ respectively. Just as in
figure \ref{fig:LeadSix2}, the labels in the diagram are determine
from the those of the first integral.} \label{fig:LeadSix1}
\end{figure}

The scalar integrals that contribute to these leading singularities
are shown in figure \ref{fig:LeadSix1}.

The location of the leading singularities is slightly more involved
as $p$ is determined as a function of $q$. Let us give the two
solutions for $q$ and then give the two $q$-dependent solutions for
$p$.
\begin{equation}
q^{(1)} = -\lambda_1\left(\tilde\lambda_1+\frac{\langle
2,3\rangle}{\langle 1,3\rangle}\tilde\lambda_2\right), \qquad
q^{(2)} = -\left(\lambda_1+\frac{[ 2,3]}{[
1,3]}\lambda_2\right)\tilde\lambda_1.
\end{equation}
The solutions for $p$ are given as $p^{(1)}(q)=(\alpha
\lambda_5+\beta \lambda_6)\tilde\lambda_q$ and $p^{(2)}(q)=\lambda_q
(\tilde\alpha \tilde\lambda_5+\tilde\beta \tilde\lambda_6)$ with
\begin{equation}
\alpha = \frac{\langle
6,4\rangle[5,6][4,q]+(s_{45}+s_{46})[5,q]}{[q,4][q|5+6|4\rangle},
\quad \beta =\frac{\langle
5,4\rangle[5,6][q,4]+(s_{45}+s_{46})[q,6]}{[q,4][q|5+6|4\rangle}
\end{equation}
and $\tilde\alpha, \tilde\beta$ the parity conjugate of $\alpha,
\beta$.

Once again we can easily write down the linear equations coming from
imposing the correct behavior at the four leading singularities. Let
us denote by $H_{p,q}$ the value of the sum over Feynman diagrams on
one of the four solutions ${p_*,q_*}$. Then the equations are
\begin{equation}
\begin{array}{ccl}
t_{456}s_{12}s_{23}(q-k_{56})^2 H_{p,q} & = & F + G(q-k_{56})^2 +
\frac{1}{(p+k_6)^2}\left(C+(E_1+D_1(p-k_1)^2)(q-k_6)^2+ \right. \\
& & \left.
(E_3+D_2(q-k_{56})^2)(p-k_1)^2+(E_4+D_3(q-k_6)^2)(p-k_{12})^2+\right.
\\ & & \left. (E_2+D_4(p-k_{12})^2)(q-k_{56})^2 \right).
\end{array}
\label{eq:second}
\end{equation}
The labels of the coefficients are explained in the caption of
figure \ref{fig:LeadSix1}.

Combining the two systems of equations, (\ref{eq:first}) and
(\ref{eq:second}), one finds eight equations.

Repeating the same procedure for the three remaining topologies in
figure \ref{fig:LeadSix3} which are completely analogous to topology
$(B)$ one finds four equations in each case. This gives a total of
twenty equations for seventeen coefficients, {\it i.e.}, eight
pentagon-box integrals and nine pentagon-pentagon integrals.

Using seventeen equations to determine the coefficients leaves three
more as consistency checks.

We end this section by mentioning that the next natural set of
equations to consider comes from double-box topologies where one of
the boxes is only a four-particle one-loop amplitude. For a given
choice of the two external legs to the four-particle amplitude, say
$\{1,2\}$, there are three topologies. These correspond to the
possible ways of distributing the remaining external legs; in our
example we would find $\{3,4,5;6\}, \{3,4;5,6\},\{3;4,5,6\}$. By
looking at the leading singularity which uses the propagator coming
form the jacobian of the four-particle box, these give rise to only
two equations each. Hence we find six equations in total. There are
however seven scalar integrals that contribute. One of them is a
hexagon-box.

Following this line of ideas it would be interesting to continue and
write down the complete set of linear equations which determines all
two-loop six particle amplitudes.

Already with the results presented here, it would be interesting to
compare the parity even part of the coefficient of all
pentagon-pentagon integrals to those obtained recently
in~\cite{Bern:2008ap} for MHV amplitudes.

\section{Conclusions}

The program of determining the S-matrix of physical theories by
studying the structure of its singularities when analytically
continued into complex values of kinematical invariants was very
ambitious. Now we know that one of the main goals of the program
which was to understand the strong interactions has proven to be a
formidable problem.

In the past 20 years we have learned that ${\cal N}=4$ super
Yang-Mills (SYM) can serve as a laboratory where new techniques and
ideas can be tested. Based on the results presented in this paper we
would like to propose that perhaps ${\cal N}=4$ SYM, at least in the
large $N$ limit, can be used to realize the basic ideas of the
$S$-matrix program. Of course, ${\cal N}=4$ SYM in four dimensions
is a conformal theory and no $S$-matrix can be defined. However, if
IR divergencies are regulated using dimensional regularization then
a sensible $S$-matrix can be defined. It is very intriguing that on
the contours which computer the discontinuity across leading
singularities, all integrals are finite and hence make perfect sense
in four dimensions. It is natural to expect the amplitudes on the
torus contours can have physical meaning. It would be interesting to
explore this further.

In general, Feynman diagrams possess many singularities; from poles,
related to resonances, to branch cuts, related to unitarity. When a
given scattering amplitude in theories with spin is considered as a
function of independent complex variables $\lambda_a^{(i)}$ and
$\tilde\lambda_{\dot a}^{(i)}$ the problem of reconstructing it from
the structure of its singularities is certainly out of hand. At
tree-level we learned few years ago
\cite{Britto:2005fq,Britto:2004ap} that one can solve a simpler
problem by concentrating on a single complex variable deformation of
the amplitude. The main simplification comes from the fact by
matching only a very small subset of all poles is enough to
determining the amplitude in terms of smaller ones. This idea led to
recursion relations between on-shell scattering amplitudes.

At loop level, Feynman diagrams posses an intricate structure of
nested branch cuts. It turns out that the discontinuity across a
given branch cut possesses branch cuts itself. The nested structure
gets more an more complicated the higher the loop order. The problem
of finding functions which reproduce all such discontinuities is
clearly very hard. Out of all these singularities, the special class
studied in this paper, called leading singularities, are the ones
which have the highest codimension. We have argued that the
discontinuities associated to them are determined by simple
computations of residues. We have shown that quite remarkably, the
problem of finding a function which reproduces all leading
singularities, which only implies the solution to linear equations,
is enough to completely determine the amplitude in ${\cal N}=4$ SYM.

Summarizing, computing multi-loop, multi-particle amplitudes in
${\cal N}=4$ SYM can be reduced to the computation of residues and
the solution of systems of linear equations. The residues only
involve products of tree-level amplitudes. These possess all the
helicity information and only enter in the inhomogeneous part of the
linear equations. The homogeneous part is universal. If we think
about the amplitude and the linear equations as being objects with
the same information, then it might be that the matrix which
determines the homogenous part of the equations is the right object
to study properties like strong coupling expansions, collinear
limits, IR consistency equations, etc.

A natural question that arises is whether this is property is unique
to ${\cal N}=4$ SYM. The most promising theories are those for which
one-loop amplitudes can be written in terms of only boxes, {\it
i.e.}, bubbles and triangles are absent. ${\cal N}=8$ supergravity
has been hypothesized to have such
property~\cite{Bern:1998sv,Bern:2005bb,BjerrumBohr:2005xx,BjerrumBohr:2006yw}.
This means that one could try to apply the technique presented here
to that case. Already, an important step in this direction was given
in \cite{Cachazo:2008dx}. It would be very interesting to compute
more examples.

In \cite{Cachazo:2008dx}, higher loop four-particle amplitudes in
${\cal N}=4$ SYM were considered. In cases where four-particle
one-loop amplitudes were present as subdiagrams it was shown that
one could related the computation to that at one less loop order. It
would be interesting to explore the consequences that matching all
leading singularities would impose on the generalized scalar
integrals. Perhaps a formal derivation of the idea of corrections
introduced in \cite{Cachazo:2008dx} can be found.

As a final note one can say that the power of matching each
individual leading singularities comes from the fact that the number
of conditions which are naturally linear grows much faster with the
number of loops than those of previous approaches.

\begin{acknowledgments}

It is a pleasure to acknowledge very useful discussions with D.
Skinner. We have also benefited from discussions with P. Benincasa,
E. Buchbinder, L. Dixon and M. Spradlin. This research is supported
by the Government of Canada through Industry Canada and by the
Province of Ontario through the Ministry of Research \& Innovation.

\end{acknowledgments}

\end{document}